\documentclass[aps,prl,twocolumn,superscriptaddress,floatfix,preprintnumbers, 10pt]{revtex4-1}

\usepackage{graphicx}
\usepackage{dcolumn}
\usepackage{bm}

\usepackage{verbatim}
\usepackage{amsfonts}
\usepackage{amssymb}
\usepackage{amsmath}
\usepackage{amsthm}
\usepackage{times}
\usepackage{xspace}
\usepackage[unicode=true,pdfusetitle,
 bookmarks=true,bookmarksnumbered=false,bookmarksopen=false,
 breaklinks=false,pdfborder={0 0 1},backref=false,colorlinks=true,citecolor=blue,
 urlcolor=blue,linkcolor=blue]{hyperref}
\usepackage{physics}

\renewcommand{\vec}[1]{\boldsymbol{#1}}

\begin{document}

\preprint{MIT-CTP-5122}

\title{Spectroscopy of spinons in Coulomb quantum spin liquids}

\author{Siddhardh C. Morampudi}
\affiliation{Department of Physics, Boston University, Boston, MA 02215, USA}
\author{Frank Wilczek}
\affiliation{Center for Theoretical Physics, MIT, Cambridge MA 02139, USA}
\affiliation{T. D. Lee Institute, Shanghai, China} 
\affiliation{Wilczek Quantum Center, Department of Physics and Astronomy, Shanghai Jiao Tong University, Shanghai 200240, China}
\affiliation{Department of Physics, Stockholm University, Stockholm Sweden}
\affiliation{Department of Physics and Origins Project, Arizona State University, Tempe AZ 25287 USA}
\author{Chris R. Laumann}
\affiliation{Department of Physics, Boston University, Boston, MA 02215, USA}


\begin{abstract}
We calculate the effect of the emergent photon on threshold production of spinons in $U(1)$ Coulomb spin liquids such as quantum spin ice.
The emergent Coulomb interaction modifies the threshold production cross-section dramatically, changing the weak turn-on expected from the density of states to an abrupt onset reflecting the basic coupling parameters.  
The slow photon typical in existing lattice models and materials suppresses the intensity at finite momentum and allows profuse Cerenkov radiation beyond a critical momentum. 
These features are broadly consistent with recent numerical and experimental results.

\end{abstract}

\maketitle

Quantum spin liquids are low temperature phases of magnetic materials in which quantum fluctuations prevent the establishment of long-range magnetic order. 
Theoretically, these phases support exotic fractionalized spin excitations (spinons) and emergent gauge fields~\cite{Anderson1987, Balents2010, Savary2017, Knolle2019}. One of the most promising candidate class of these phases are U(1) Coulomb quantum spin liquids such as quantum spin ice - these are expected to realize an emergent quantum electrodynamics~\cite{Motrunich2002, Wen2003, Huse2003, Moessner2003, Hermele2004, Castelnovo2012, Gingras2014}. Establishing the exotic phenomena in this context would provide a foundation for exploring other conjectured phases of matter.  It will also allow us to explore regimes of quantum electrodynamics which are theoretically interesting, but otherwise inaccessible. 

At present, the main method to diagnose a spin liquid experimentally is through the the absence of distinct features associated with a local order parameter such as Bragg peaks, and instead the presence of a broad continuum in neutron scattering indicative of a multi-particle continuum. However, broad continua can also arise from other causes and one would like to have more specific signatures which highlight the emergent gauge field.  Here we identify and study features in the zero-temperature cross-section for spinon production in Coulomb quantum spin liquids which directly reflect central aspects of the underlying theory, including the existence and the unusual nature of the emergent photon. 

 \paragraph{Threshold behavior---}

The emergent photon in quantum spin ice arises from coherent ring-exchange processes which lift massive degeneracy within the manifold of spin configurations consistent with classical ice rules. 
As ring-exchange is typically a weak process, the photon propagates with a small speed $c$ and has a small bandwidth set by the Brillouin zone cutoff.
On the other hand, the spinons typically propagate due to direct spin exchange interactions, which can be parametrically larger than the ring exchange process.
This contrast leads to a strongly non-relativistic theory in which spinons readily propagate faster than the emergent speed of light. 
Taking into account both the gapless nature of the photon and its slow speed leads us to predict distinctive features in the cross section.

\begin{figure}[b]
    \centering
    \includegraphics[width=\columnwidth]{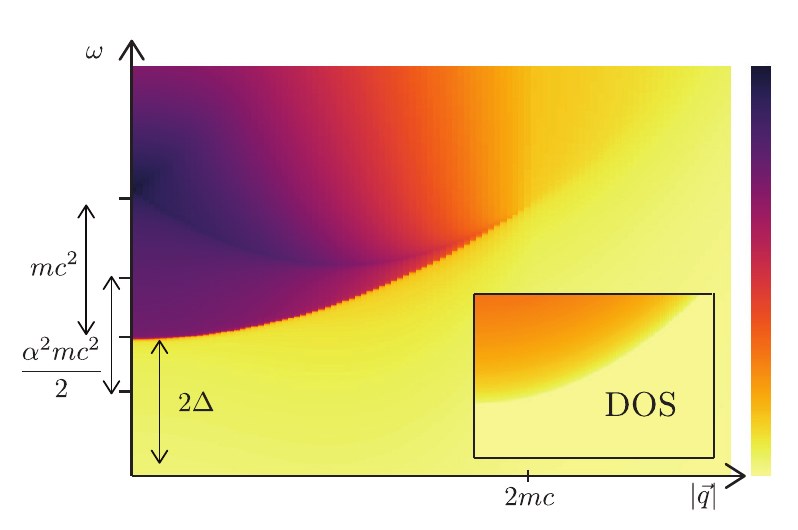}
    \caption{Dynamic structure factor $S(q, \omega)$ in a Coulomb quantum spin liquid measuring production of two spinons near threshold. 
    Compared to the naive density of states (inset), the threshold intensity is strongly enhanced for small $q$ due to the  emergent Coulomb interaction over an energy range of order the Rydberg scale $\alpha^2mc^2/4$ above threshold $2\Delta$.
    Bound Rydberg states (not shown) accumulate at the Rydberg scale below the threshold. 
    Breit interactions due to the transverse photon reduce the threshold enhancement with increasing $q$. 
    With larger momentum, the spinons exceed the emergent speed of light and emit Cerenkov radiation. 
    This causes the threshold to disappear into a diffuse continuum for $q>2mc$ and a peak in intensity for $\omega \sim (\frac{q}{2} - mc)^2/m$.
    Both plots use the same scales.
    }
    \label{fig:contour}
\end{figure}

The most dramatic consequence of the Coulomb interaction between the spinons is a  universal non-perturbative enhancement of the threshold cross section for spinon pair production at small momentum $\vec{q}$. 
In this regime, the dynamic structure factor in the spin-flip sector observed in neutron scattering exhibits a step discontinuity,
\begin{align}
\label{eq:jumpintensity}
    S(\vec{q}, \omega) &\sim S_0 \left(1 - \frac{1}{4}\left(\frac{q}{mc}\right)^2 \right) \theta(\omega - 2\Delta - \frac{q^2}{4m})
\end{align}
rather than the naive square root onset predicted by the density of states for spinon pairs~%
\footnote{The factor of $1/4$ in Eq.~\eqref{eq:jumpintensity} follows from an approximate semi-classical analysis of the Breit Hamiltonian below. We expect the correct asymptotic factor to be no larger than $1/4$ and no smaller than $1/8$.}. 
Here, $m$ and $\Delta$ are the effective mass and gap for the spinons and $c$ is the emergent speed of light.
The threshold intensity jump, $S_0 \propto m^2 e^2 = m^{2} c \alpha$, is proportional to the strength of the Coulomb interaction and provides a measure of the emergent fine structure constant $\alpha$. 

The strong onset in Eqn.\,(\ref{eq:jumpintensity}) is analogous to the Sommerfeld enhancement~\cite{Sommerfeld1931,Landau1981} observed in semiconductor exciton production~\cite{YuCardona2010}.  However, here the small speed of light means that transverse photon exchange has non-negligible consequences.
Indeed, the transverse interaction is responsible for the suppression of the enhancement with $\left(\frac{q}{mc}\right)^2$ at finite momentum $\vec{q}$. 
More dramatically, since spinons emits Cerenkov radiation when their velocity exceeds the speed of light, there is a finite lifetime for spinons propagating at high energy and momenta.  For momenta $\vec{q} > 2mc$, even threshold spinons have a finite Cerenkov lifetime, and then the threshold in the dynamic structure factor becomes entirely diffuse. 

The universality of the threshold behavior follows from Wigner's insight, according to which the energy dependence of cross sections just above threshold are governed by long distance interactions between the slowly escaping particles\cite{Wigner1948, Morampudi2017}. 
In our case, these are the Coulomb and Breit interactions expressed in Eq.~\eqref{eq:Ham}, which can be analyzed semiclassically in the long distance region. 
The short distance scattering wavefunctions are, of course, sensitive to lattice scale effects, but generically those vary smoothly with energy near threshold.
Thus, the jump at small $\vec{q}$ and associated low energy spectral weight for spinon production are remarkably direct signatures of the emergent gauge theory.

\paragraph{Computation---}
A minimal model for the spinon dynamics is given in an effective mass approximation by the following Lagrangian,
\begin{align}
\label{eq:lagrangian}
    \mathcal{L} &=  \psi_\sigma^{\dagger}(i\partial_{t}-\sigma e\phi )\psi_\sigma - \frac{|(-i\nabla - \sigma \frac{e}{c} \vec{A})\psi_\sigma|^2}{2m}  \nonumber \\
    & - \Delta|\psi_\sigma|^2 - V(\psi)
\end{align}
where $\psi_\sigma$ represent the spinon ($\sigma=+1$) and anti-spinon ($\sigma = -1$) fields, $e$ is the emergent charge, $m$ is the effective mass and $\Delta$ is the spinon gap. 
The higher-order interaction potential $V(\psi)$ contains all of the short-range interactions between the spinons including those induced by gapped, weakly dispersive visons (magnetic monopoles), which we do not otherwise attempt to model.
The emergent scalar $\phi$ and vector $\vec{A}$ potentials are governed by the usual Maxwell Lagrangian (in CGS units),
\begin{align}
    \mathcal{L}_{EM} &= 
    \frac{1}{8\pi}\left[ (\nabla\phi)^2 + \frac{1}{c^2}(\partial_{t} \vec{A})^2 - (\nabla \times \vec{A})^2\right] 
\end{align}
with emergent speed of light $c$. 
We work in Coulomb gauge $\vec{\nabla}\cdot\vec{A} = 0$ throughout.
(Note that if the system possesses an additional global $U(1)$ spin symmetry, as in the XXZ model for quantum spin ice, then the spinon fields must be doubled to account for the fractional assignment of the global charge. 
That does not change the results presented here qualitatively.)

The neutron scattering cross-section is proportional to the dynamic structure factor $S(\vec{q},\omega)$, given by the imaginary part of the dynamic spin susceptibility $\chi(\vec{q}, \omega)$. 
The interesting part comes from pair production of spinons. Neglecting all interactions, this is given by the bubble diagram 
\begin{align}
    \chi^{(0)} (\vec{q}, i\omega) = \vcenter{\hbox{\includegraphics[scale=0.5]{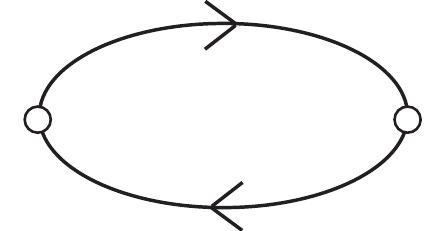}}}
\end{align}
At zero temperature, this produces the usual 3D density of states, $S^0(\vec{q}, \omega) \sim m^{3/2} \sqrt{\omega - 2\Delta - q^2/4m}$. 

The effects of interactions can be taken into account with renormalized propagators $G(\vec{k}, i\kappa)$ (double lines) and a vertex $\Gamma(\vec{k}_1, \vec{k}_2, i \kappa_1, i \kappa_2)$ (triangle).
\begin{align}
\label{eq:chibubfull}
    \chi (\vec{q}, i\omega) = \vcenter{\hbox{\includegraphics[scale=1]{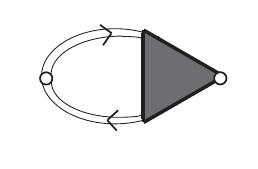}}}
\end{align}
where the vertex $\Gamma$ is defined through an irreducible 2-particle diagram (hatched square)
\begin{align}
\label{eq:bethesalpeter}
 \vcenter{\hbox{\includegraphics[scale=1]{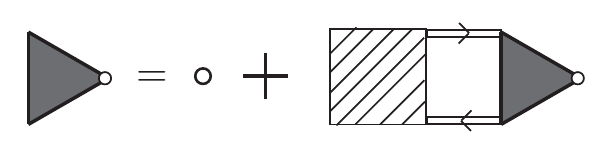}}}
\end{align}

We work with a renormalized mass $m$ which takes into account the effect of interactions in $G(\vec{k}, i\kappa) \equiv \dfrac{1}{i\kappa - k^2/2m - i\Im\Sigma(\vec{k}, i\kappa)}$ and neglect higher-order corrections to the dispersion. We discuss lifetime effects in the next section and suppress the self-energy $\Sigma$ until then. All the momenta have a UV cut-off due to the Brillouin zone. There are higher order corrections to the photon dispersion on a lattice, but they will not affect the effects described below.

Among diagrams resulting from interactions, crossed diagrams in the two-particle irreducible vertex $\Gamma$ can be neglected (ladder approximation) since they are suppressed by products of the Bose occupation factor $n_B$ which vanish at temperature well below the spinon gap $2\Delta$. 
Thus, the dominant diagrams in the perturbative expansion are ladder diagrams with Coulomb interactions (dashed line), transverse photon exchange (wavy line) and short-range interactions (line with star) between the spinon and anti-spinon, 
\begin{align}
    \vcenter{\hbox{\includegraphics[scale=1]{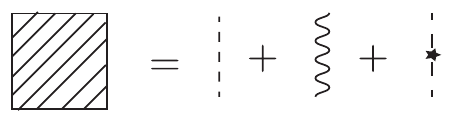}}}
\end{align}

Just above threshold, we can neglect the frequency dependence of the photon propagator as the spinons separate asymptotically slowly. 
In this approximation, the vertex $\Gamma$ is independent of the relative frequency between the two spinons, and Eq.~\eqref{eq:chibubfull} can be reduced to $\chi(\vec{q}, i\omega) = W(0; \vec{q},i\omega)$ where
\begin{equation}
W(\vec{r}; \vec{q}, i\omega) \equiv \sum_{\vec{k}}\dfrac{e^{i \vec{k}.\vec{r}} \Gamma(\vec{q}/2 + \vec{k}, \vec{q}/2 - \vec{k},  i\omega)}{i\omega - k^2/m- q^2/4m -2\Delta}
\end{equation}
We have introduced $W$ for convenience and temporarily neglected radiation effects in the propagators.

Using the Bethe-Salpeter equation (Eq.~\eqref{eq:bethesalpeter}) for the vertex, we see that $W(\vec{r}; \vec{q}, i\omega)$ is the Green's function for the Schrodinger equation governing the relative spinon motion
\begin{align}
    (i \omega - H_s - 2\Delta) W(\vec{r}; \vec{q}, i\omega) = \delta^{3} (r)
\end{align}
where 
\begin{align}
\label{eq:Ham}
    H_s &\equiv \dfrac{\hat{q}^2}{4m} + \dfrac{\hat{p}^2}{m} - \dfrac{ e^2}{r}+ V(\vec{r}) \nonumber \\
    &+ \dfrac{e^2}{2m^2c^2r}\left(\frac{\hat{q}^2}{4} + \dfrac{(\hat{\vec{q}}.\vec{r})^2}{4r^2} - \hat{p}^2 - \dfrac{\vec{r}\cdot(\vec{r}\cdot \hat{\vec{p}}) \hat{\vec{p}}}{r^2}\right)
\end{align}
Here, $\hat{\vec{q}}$ and $\hat{\vec{p}}$ correspond to the center-of-mass and relative momentum. The terms suppressed by $c$ are Breit terms governing the leading velocity dependent interactions between moving charges. 
Using a spectral representation for $W$, we find
\begin{align}
\label{eq:fgr}
    S(\vec{q}, \omega) = 2\pi\sum_{j} |\psi_j(r=0)|^2 \delta(\omega - 2\Delta - \frac{q^2}{4m} - \epsilon_j)
\end{align}
where $\psi_j$ are the relative eigenstates of $H_s$. 
Physically, Eq.~\eqref{eq:fgr} shows that the naive density of states is enhanced by the probability that the two spinons be found together in the eigenfunction.

The preceding derivation of the effective two-body Schrodinger equation holds for energy close to threshold and small spinon momenta. 
In particular, the Breit term, corresponding to instantaneous current-current interactions between the spinons, is only valid when the spinons are moving sufficiently slowly and thus should not be used directly in the short distance region  \cite{Breit1932, BetheSalpeter2012}. 
Fortunately, the energy dependence of the probability at the origin $|\psi_j(0)|^2$ is governed by the long range part of the interactions near the threshold. 
The short distance wavefunction, whatever it is, is rigid due to the large energy costs associated with amplitude shifts in that region \cite{Wigner1948, Morampudi2017}. We can thus neglect both $V(\vec{r})$ and the exact form of the Breit Hamiltonian for small $r$ in estimating the energy dependence of $|\psi_j(0)|^2$.

At large separation, the effective potential between the spinons decays as $1/r$ with a $\vec{q}$-dependent effective charge $e\rightarrow e\sqrt{1-\dfrac{q^2}{4m^2c^2}}$ restricted to the zero angular momentum channel. 
A semi-classical analysis of the mixed terms involving products of $\vec{r}$ and $\vec{p}$ (which are neither pure potential nor kinetic) shows that they only change the probability  $|\psi_j(0)|^2$ by an energy-independent constant.

Using the renormalized effective charge from the Breit interactions and the solution of the two particle Coulomb problem, we find
\begin{align}
    S(\vec{q}, \omega) \sim \dfrac{m^{3/2}\sqrt{2\pi R}}{1-\exp(-\sqrt{\frac{2\pi R}{\omega - 2\Delta - \frac{q^2}{4m}}})}\theta(\omega - 2\Delta - \frac{q^2}{4m})
    \label{eq:bottom}
\end{align}
where $R=\frac{1}{4}mc^2\alpha^2(1 - \frac{q^2}{4m^2c^2})$ is the effective Rydberg constant. Here, $\alpha=e^2/c$ is the emergent fine-structure constant.
Neglecting the (non-universal) essential singularity cutting off the enhancement for energies above $R$ recovers Eq.~\eqref{eq:jumpintensity}.

The two particle problem also has an infinite set of bound state solutions (excitons) at energies $\omega_n = 2\Delta - \frac{R}{n^2}$ below the two particle continuum.  The $q$-dependent charge renormalization implies that the states bend into  the two-particle continuum as $q$ approaches $2mc$. 
A similar pattern of bound states is seen in studying defects in quantum dipolar spin ice from a mapping to a Coulomb problem on the Bethe lattice \cite{Petrova2015}. 
We have not shown bound states in Fig.~\ref{fig:contour} since their lifetime is affected by temperature, disorder, and other complicating effects. 

\paragraph{Cerenkov radiation---}
Since the system is not Lorentz invariant, the spinons can exceed the effective speed of light.
The resulting Cerenkov radiation carries momentum and energy away from the spinon and leads formally to a finite lifetime (imaginary part of the self-energy) for $k > mc$. 
The leading order contribution to the spinon lifetime comes from
\begin{align}
\label{eq:cerenkovdiagram}
    \Sigma(\vec{k}, i\omega) &\sim \vcenter{\hbox{\includegraphics[scale=0.5]{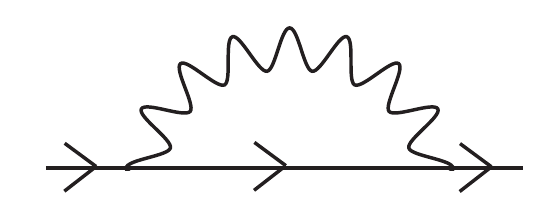}}}
\end{align}
The seagull diagram, although of the same order in $\alpha$, does not contribute to the lifetime. 
There are no self-energy corrections to the photon for temperature  well below the spinon gap $\Delta$. 

After analytic continuation, the imaginary part of Eq.~\eqref{eq:cerenkovdiagram} is given by
\begin{align}
    \Im \Sigma (\vec{k}, \omega) &\sim \frac{e^2 k^2}{mc} \int_{0}^{\theta_c} d\theta \int_{0}^{k_{\Lambda}} dK K \sin^{3} (\theta)\nonumber\\ 
    &\times \delta(\omega - \frac{(\vec{k} - \vec{K})^2}{2m}- cK)
\end{align}
where $\theta$ is the angle between $\vec{k}$ and the photon momentum $\vec{K}$, and $\theta_c$ is the critical angle below which no photons are radiated (ie. the Cerenkov cone). The finite bandwidth $k_{\Lambda}$ can cut off the amount of radiation produced if $k_{\Lambda} < 2(k - mc)$. Assuming no such cut off,
\begin{align}
    \Im \Sigma (\vec{k}, \omega) &\sim  \frac{e^2k^2}{mc}\int_{0}^{\theta_c} d\theta \sin^{3} (\theta) \\ &\times (1 + \frac{k\cos (\theta) - mc}{\sqrt{(k\cos (\theta) - mc)^2 + 2m\omega - k^2}}) \nonumber
\end{align}
If we evaluate the self-energy on-shell, $\cos (\theta_c) = mc/k$ and we get
\begin{align}
    \Im \Sigma (\vec{k}, \omega=\frac{k^2}{2m}) &\sim \frac{e^2 k^2}{mc}(\frac{mc}{k} - \frac{1}{3}(\frac{mc}{k})^3 - \frac{2}{3})
\end{align}
when $k\geq mc$. Close to the threshold for radiation, the inverse lifetime goes as $e^2 (k - mc)^2/mc$, whereas for $k>>mc$, it goes as $2e^2k^2/3mc$. 

At threshold, the spinons have no relative momentum, and both start emitting Cerenkov radiation at a critical external momentum transfer $q=2mc$ which gives each of them momentum of $mc$ and velocity $c$. Since the relative momentum is zero, the vertex equation can be again solved including the self-energy effects. The resulting effect is that at momenta beyond $2mc$, the threshold becomes increasingly diffuse (Fig.~\ref{fig:contour}).

\begin{figure}[t]
    \centering
    \includegraphics[width=\columnwidth]{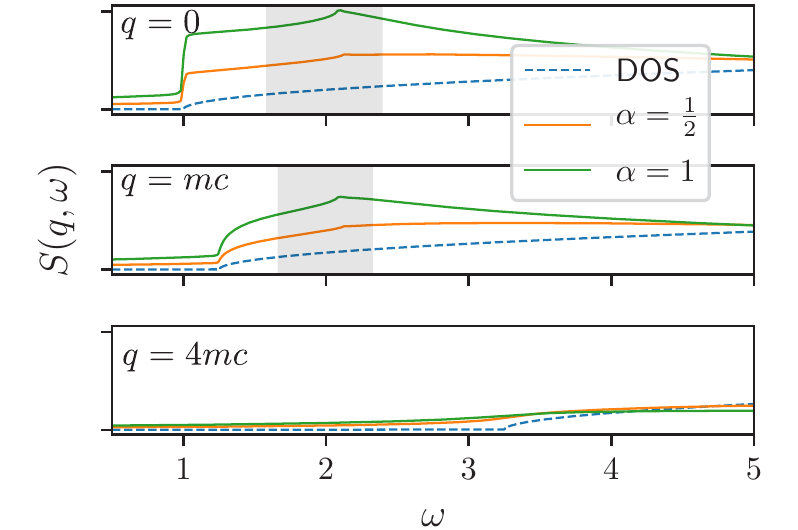}
    \caption{$S(q, \omega)$ (arbitrary units) at varying momentum $q$ ($\Delta$ = 0.5, $m=c=1$).  The naive density of states (dashed blue) shows a square-root onset.  Interaction with the gauge field enhances the square root into a jump discontinuity  with magnitude proportional to the emergent fine-structure constant $\alpha=e^2/c$. 
    The size of the jump decreases with $q$ due to the Breit interaction from  transverse photon exchange (second panel), until the threshold is washed out entirely by the production of Cerenkov radiation at all energies (third panel).
    The peaks in the grayed region arise from spectral transfer from the high energy regime, where the spinons are broadened significantly by radiation toward the low energy region where no radiation is produced. 
    The exact shape of the peaks in the grayed regime is likely an artifact of the approximations used, which are most accurate close to threshold.
    }
    \label{fig:lineCuts}
\end{figure}

The spinons also emit Cerenkov radiation at any external momentum once they have enough energy above threshold, i.e., when $\omega > \omega_c(q) = (\frac{q}{2} - mc)^2/m$. The vertex equation cannot be solved exactly in this regime, but the dominant effects are captured by considering collinear trajectories of the spinons in the evaluation of the self-energies. Additionally, retardation effects from the transverse photon will be important away from threshold and will contribute to a cut-off of the Sommerfeld enhancement at relativistic speeds. The qualitative effect is a peak in intensity when $\omega \sim \omega_c(q)$ due to a spectral weight transfer (Fig.~\ref{fig:lineCuts}). 

\paragraph{Top threshold---}
Near the top of the band, the effective mass is negative. The dominant vertex corrections at the top threshold can be taken into account by noticing that the problem can be mapped to a positive mass spinon and anti-spinon interacting with an repulsive Coulomb interaction. This suppresses the intensity at the top threshold and gives\footnote{In this section, all the parameters such as $\Delta$ and $m$ refer to an effective mass approximation around the top of the band.}
\begin{align}
S(\vec{q}, \omega) \sim m^{3/2}\sqrt{2\pi R}\exp(-\sqrt{\frac{2\pi R}{2\Delta - \frac{q^2}{4m} - \omega}})
\end{align}
However, kinematically the negative mass allows for Cerenkov radiation right at threshold. This means that near the top of the band the threshold is diffuse for all momentum transfers. The combined effect of the suppression and Cerenkov radiation at threshold implies that the spectral intensity at the top is heavily suppressed compared to the density of states.

\paragraph{Application to spin ice---}
The results apply to any U(1) Coulomb quantum spin liquid with gapped spinons, such as have been discussed in models of hard-core bosons on the pyrochlore \cite{Motrunich2002}, interacting dipoles \cite{Wen2003}, quantum rotors \cite{Wen2004} and quantum link models \cite{Wiese2013}. Perhaps the most promising experimental application is to quantum spin ice materials \cite{Ross2011, Savary2012, Kimura2013, Gingras2014, Glaetzle2014, Kato2015, Anand2016, Petit2016, Wan2016, Tokiwa2018}. The ideal realization is an XXZ model on the pyrochlore lattice\cite{Huse2003, Hermele2004, Banerjee2008, Shannon2012, Benton2012} with parameters $J_{zz},J_{\pm}$ such that $J_{\pm} \ll J_{zz}$, and the generic Hamiltonian describing spin ice materials contains perturbations to this. The spinons live on a diamond lattice with the bare mass given by $m \sim \hbar^2/4J_{\pm}a_0^2$ and a gap $\Delta \sim J_{zz}/2 - 12J_{\pm}$. The photon has a bandwidth set by the ring-exchange $g\equiv 12J_{\pm}^3/J_{zz}^2$ and hence a speed of light $c\sim \xi g a_0/\hbar$ where $\xi$ is an $O(1)$ constant and $a_0$ is the lattice constant. There would be a significant enhancement in intensity seen in neutron scattering over an energy scale $R =  \dfrac{\alpha^2 m c^2}{4} \sim \dfrac{9\xi^2 \alpha^2 J_{\pm}^5}{J_{zz}^4}$. Additionally, the threshold will become incoherent due to Cerenkov effects at $q_c = 2mc \sim \dfrac{3\xi\hbar J_{\pm}^2}{a_0 J_{zz}^2}$. A more accurate estimate would use the renormalized parameters determined from data as detailed in the discussion.

Recent numerical and experimental works have made progress in determining the spinon dynamic structure factor in quantum spin ice. Quantum Monte Carlo data from \cite{Huang2018} clearly shows a sudden onset in intensity at the bottom threshold and a drop in intensity at threshold at large wavevectors away from the bottom, indicative of possible Cerenkov effects. Neutron scattering data on $\mathrm{Pr}_2\mathrm{Hf}_2\mathrm{O}_7$, a candidate quantum spin ice, also shows a jump in intensity at the expected threshold for spinon excitations as opposed to a slow turn expected from density of states \cite{Sibille2018}. This is consistent with our predictions, but more precise data is needed to confirm this and extract quantitative parameters such as the fine-structure constant. Ref~\cite{Udagawa2019} also shows a jump in intensity at threshold in the momentum-integrated structure factor at intermediate temperatures. However, we do not believe this is related to the emergent gauge field interactions described here since it requires a choice of the sign of $J_{\pm}$ which produces a fine-tuned van Hove singularity in the spinon dispersion at threshold.

\paragraph{Discussion---}
We can use predictions from the effective theory to determine its fully renormalized parameters, using numerical or experimental data on spinon cross-sections. The mass and gap can be determined from the onset and curvature of the threshold respectively. The enhancement of the response due to the photon makes this measurement easier than the case where the response only corresponds to density of states which vanishes at threshold. The speed of light can then be determined as $c=q_c/2m$ where $q_c$ is the momentum at which the threshold starts to become diffuse.

The fine-structure constant can also be determined from the threshold behaviour. It determines the magnitude of the jump discontinuity. Since the overall intensity is multiplied by non-universal prefactors, to extract the effective Rydberg $R$ and corresponding fine-structure constant using Eq.~(\ref{eq:bottom}) it is necessary to consider the energy and/or momentum dependence of the intensity.

In conclusion, let us note that the general possibility to find emergent gauge structure in ``mechanical'' models has a glorious history, as it led Maxwell to his equations.  Today, given our increased abilities to engineer interactions and to sculpt metamaterials (such as spin-ice physics at room temperature\cite{Coates2019}), it is newly relevant to practice, even if it does not lead to new models of microphysics. Emergent electrodynamics can give us access to parameter regimes and phenomena that are not easily accessible otherwise, e.g. magnetic monopoles and dyons, the regime of slow light and copious Cerenkov radiation, and strong coupling.  It can also be interesting as a testbed for approximation schemes or, more speculatively, as a tool for quantum simulation.

\begin{acknowledgements}
\section*{Acknowledgments}
The authors would like to thank Anushya Chandran, Bruce Gaulin, Patrick Lee, Roderich Moessner, Boris Spivak and Senthil Todadri for useful discussions.
FW's work is supported by the U.S. Department of Energy under grant Contract  Number DE-SC0012567, by the European 
Research Council under grant 742104, and by the Swedish Research Council under Contract No. 335-2014-7424.
CRL acknowledges support from the NSF through grant PHY-1752727.
\end{acknowledgements}

\bibliography{refs}

\end{document}